\documentclass{elsart}

\usepackage{graphicx}

\usepackage{amssymb}

\begin{document}

\begin{frontmatter}



\title{Development of Gaseous Tracking Devices for the Search of WIMPs}


\author[KYOTO]{H. Sekiya\corauthref{cor}},
\corauth[cor]{Corresponding author. tel:+81 75 753 3868; fax:+81 75 753 3799.}
\ead{sekiya@cr.scphys.kyoto-u.ac.jp}
\author[KYOTO]{K. Hattori},
\author[KYOTO]{S. Kabuki},
\author[KYOTO]{H. Kubo},
\author[KYOTO]{K. Miuchi},
\author[WASEDA]{T. Nagayoshi},
\author[KYOTO]{H. Nishimura},
\author[KYOTO]{Y. Okada},
\author[KOBE]{R. Orito},
\author[KYOTO]{A. Takada},
\author[ICRR]{A. Takeda},
\author[KYOTO]{T. Tanimori},
\author[KYOTO]{K. Ueno}

\address[KYOTO]{Department of Physics, Graduate School of Science, Kyoto University, Kitashirakawa, Sakyo, Kyoto, 606-8502, Japan}

\address[WASEDA]{Advanced Research Institute for Science and
 Engineering, Waseda University, \\
17 Kikui-cho, Shinjuku, Tokyo, 162-0044, Japan}

\address[KOBE]{Department of Physics, Graduate School of Science and Technology, Kobe University, 1-1 Rokkoudai, Nada, Kobe, 657-8501, Japan}

\address[ICRR]{Kamioka Observatory, ICRR, University of Tokyo,\\
 456 Higasi-mozumi, Hida-shi, Gifu, 506-1205, Japan}

\begin{abstract}
The Time Projection Chamber (TPC) has been recognized as a potentially 
powerful detector for the search of WIMPs by measuring the directions
of nuclear recoils, in which the most convincing signature of WIMPs, 
caused by the Earth's motion around the Galaxy, appears. 

We report on the first results of a performance study of the
neutron exposure of our prototype micro TPC  
with Ar-C$_2$H$_6$ (90:10) and CF$_4$ gas at 150 Torr.

\end{abstract}

\end{frontmatter}

\section{Introduction}
\label{intro}
It is considered by many that the galactic halo is composed of weakly
interacting massive particles (WIMPs) as dark matter\cite{jung}. 
These particles could be detected directly by measuring the nuclear
recoils produced by their elastic scattering off nuclei in detectors.
The most convincing signature of WIMPs appears in the directions of
nuclear recoils. It is provided by the Earth's large velocity through
the isothermal galactic halo ($\sim$230 km/s). Hence, detectors sensitive to 
the direction of the recoil nucleus would have a great potential to
identify WIMPs\cite{dir}. 

Time Projection Chambers (TPCs) with fine spacial resolutions are among
such devices, and we are developing a micro TPC, which can detect 
three-dimensional fine tracks of charged particles\cite{miuchi}.
Since the energy deposits of WIMPs to nuclei are only a few tens of keV
and the range of nuclei is limited, the micro-TPC should be operated 
at low pressures.
 
We also focused on the detection of WIMPs via spin-dependent(SD) interactions 
and are interested in operating the micro-TPC with CF$_4$\cite{NA},
because $^{19}$F has a special sensitivity to 
SD interactions for its unique spin structure\cite{collar}.

In the present work, in order to examine the response of the 
micro-TPC to nuclear recoils at low pressures as a first step,
we irradiated a 150 Torr Ar-C$_2$H$_6$ (90:10 mixture) 
gas (one of the standard gases for TPCs) and CF$_4$ with neutrons 
from $^{252}$Cf.
The track lengths and deposited energies
of Ar, C, and F recoils were investigated.

\section{The micro-TPC}
\begin{figure}[p]
\begin{center}
\includegraphics[width=10cm]{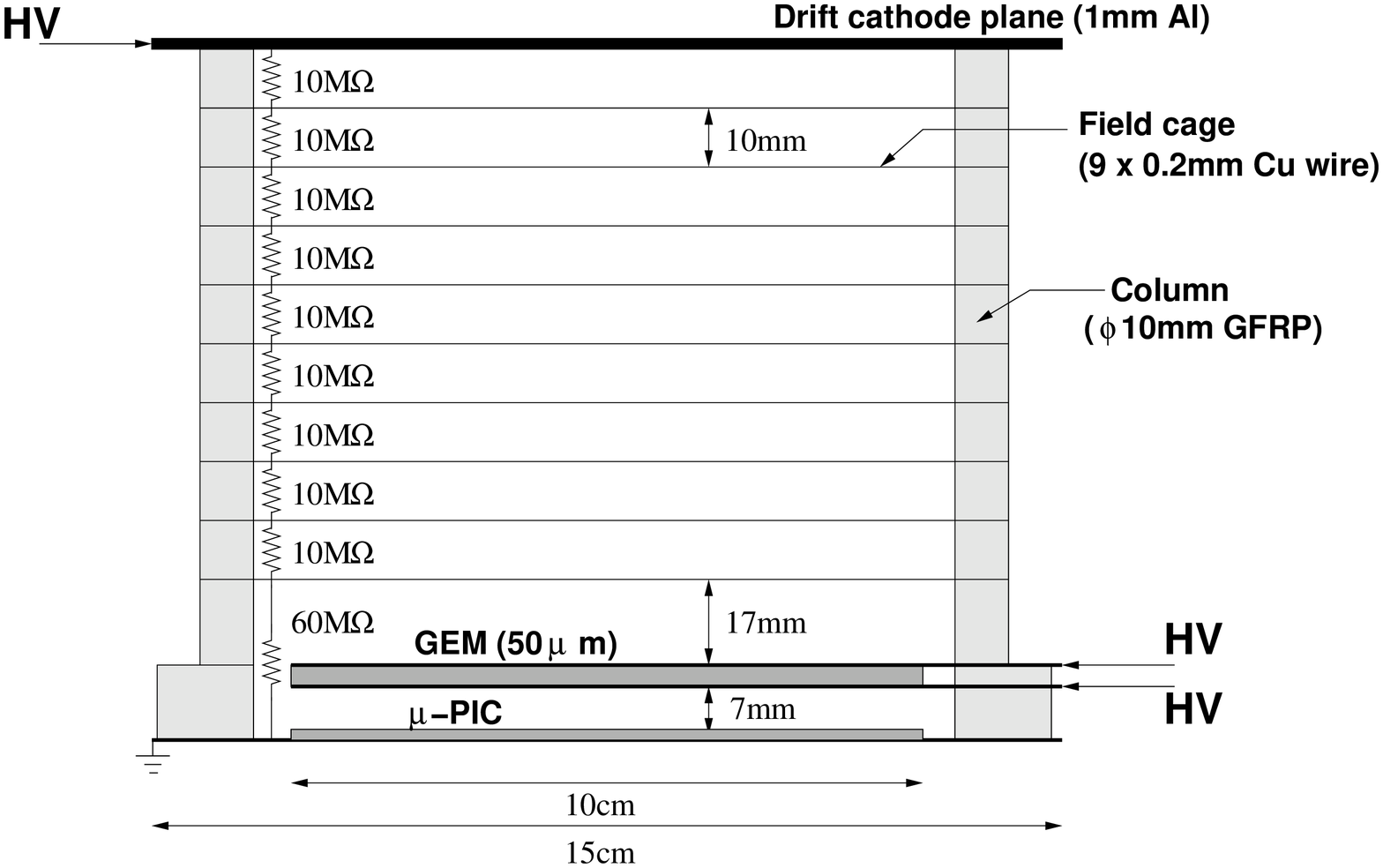}
\caption{Schematic diagram of the prototype micro-TPC and the 
drift-field cage.}
\label{fig:DC}
\end{center}
\end{figure}

The prototype micro-TPC used in this measurements 
is shown in Fig.\ref{fig:DC}.
The field cage consists of a drift cathode plane
and nine 0.2 $\mu$m copper wires of 1cm pitch with connections 
of 10 M$\Omega$ resistor, which forms a uniform electric field
in the detection volume of $10\times10\times10$ cm$^3$. 

The $\mu$-PIC\cite{upic} for 2-dimensional readout is $10\times10$cm$^2$ 
with 256 anode strips and 256 cathode strips each with a 400 $\mu$m pitch.
We also used a GEM having a 10 cm$\times$10 cm$^2$ sensitive area as a
sub-amplification device between the field cage and $\mu$-PIC, as
illustrated in Fig. \ref{fig:DC}, which enables stable operation 
and avoids discharges with low HV operation of 
both the $\mu$-PIC and GEM. 
The details of this GEM are described in Refs.\cite{gem,hattori}.
  
The output charges of $256+256$ channels
are pre-amplified (0.7 V/pC) and shaped (with a gain of 7) and 
discriminated via ASD chips (4 channels/chip, SONY CXA3653Q)\cite{asd}.
The pre-amplified signals are summed 
and digitized by 100 MHz 8bit flash ADCs
in order to determine the deposited energy and the 
track direction as the waveforms hold the Bragg curve shapes.  

The reference threshold voltage ($0-100$mV) is commonly supplied to
all the ASD chips and all discriminated digital signals are sent to 
 the position encoding module based on FPGAs with an internal
clock of 100 MHz, so that the anode and cathode coincident position (x,y)
 and the timing (z) are recorded in the memory module and the
tracks of charged particles are reconstructed in software.  
The tracking performances for electrons, protons, and MIPs 
are reported elsewhere\cite{miuchi,hattori}.

\section{Measurements and Results}
\begin{figure}[p]
\begin{center}
\includegraphics[width=10cm]{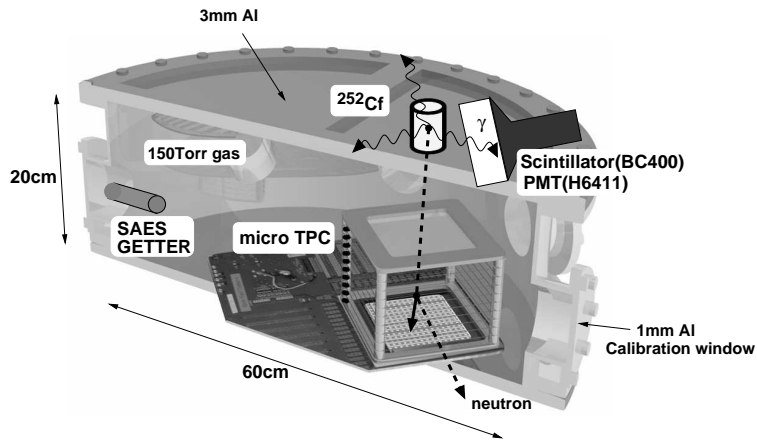}
\caption{Setup of this measurement for 150 Torr gas operation.}
\label{fig:Setup}
\end{center}
\end{figure}

As illustrated in Fig. \ref{fig:Setup},
the micro-TPC was set in a 6 mm-thick aluminum vessel of 
60 cm diameter $\times$ 20cm height.
In a typical run,
the vessel was evacuated to $\sim8\times10^{-3}$ Torr,
the SAES GETTER$^{\mbox{\scriptsize{\textcircled{\tiny R}}}}$ pump
in the vessel was activated,
and then the vessel was filled with Ar-C$_2$H$_6$ (90:10) or CF$_4$ gas   
to a pressure of 150 Torr and sealed for the duration of the measurement.

For measuring the gas gain and the energy calibration,
the gas was irradiated with 
$^{109}$Cd 22 keV and $^{133}$Ba 31.0 keV X-rays 
through a 1mm thick aluminum window close to the sensitive volume.

We irradiated the micro-TPC with neutrons from a 1 MBq $^{252}$Cf source
on the top of the vessel.
Since one fission decay of $^{252}$Cf emits 3.8 neutrons and 9.7 
$\gamma$-rays on average\cite{neut}, the $\gamma$-rays or neutrons  
detected by a $10\times10\times2$ cm$^3$ plastic scintillator 
were used as the event trigger.

In the $\gamma$/n-triggered events, gamma events would dominate
under normal gas gain ($\sim$10000) operation. Since  
the $dE/dx$ values of the neutron events are much larger than 
those of gamma events, we operated the $\mu$-PIC and GEM 
with a rather low gas gain (below 1000) in order to observe the nuclear recoils.

In such different gas gain measurements,
we fixed the anode voltage of the $\mu$-PIC and changed the voltage 
between the GEM electrodes. 
Below a gas gain of about 2000, our system was not able to measure the 
 $^{109}$Cd 22 keV x-ray correctly due to a mismatch of the dynamic 
range of the ASD chips and the flash ADC;
therefore, the deposited energy in low-gain operations was extrapolated
from the calibrations with the high gas gain operations. 

We evaluated the track length
as a function of the measured electron equivalent energy in the
following way.

\subsection{Ar-C$_2$H$_6$ 150Torr run}
The drift cathode plane was supplied $-1$ kV,  which gave a drift field of
60 V/cm and an electron drift speed of 4.0 cm/$\mu$s.
The anode voltage of the $\mu$-PIC was fixed at 350 V.

For nuclear recoil measurements,
the threshold of the discriminator of the ASD chip was set to 80 mV and 
the measured track length of events  
when the GEM voltage was set to 200 V (gas gain of 3000) and 135 V 
(gas gain of 900) is shown in Fig. \ref{fig:ArTE}.
The MC (Geant4\cite{geant}) simulated track length without
consideration of the diffusion, the energy resolution, and 
the $dE/dx$ threshold is also indicated for a comparison.
The geometry used for the simulation was in accordance with
Fig. \ref{fig:DC} and Fig. \ref{fig:Setup} and, the neutron energy 
spectrum of the spontaneous fissions of $^{252}$Cf was assumed to be
\begin{equation}
\frac{dN}{dE}=\sqrt{E}\exp\left(-\frac{E}{T}\right), 
\end{equation}
where $T=1.3$ MeV\cite{fission}.

Under operation with a gas gain of  3000,  
electron recoils and proton (of C$_2$H$_6$) recoils were 
clearly observed  according to their $dE/dx$.
On the other hand, in the operation of the gas gain of 900, 
the C and Ar recoils and some proton recoils were observed due to
the high $dE/dx$ threshold.    


\subsection{CF$_4$ 150Torr run}
The drift cathode plane was supplied $-2$ kV,  which gives a drift field of
120 V/cm and the electron drift speed of 12.0 cm/$\mu$s.
The anode voltage of the $\mu$-PIC was fixed at 600 V.

\begin{figure}[p]
\begin{center}
\includegraphics[width=7cm]{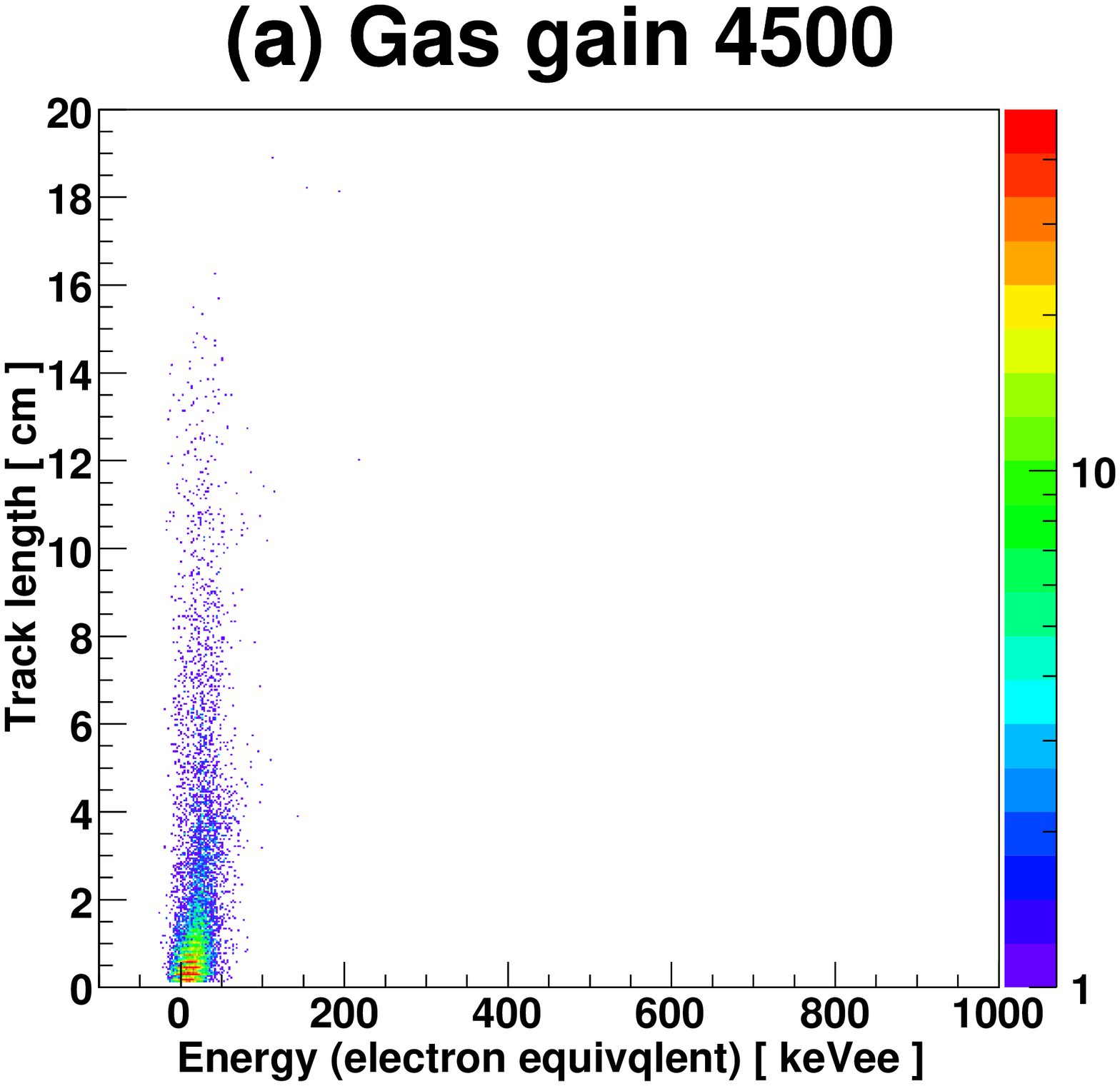}
\includegraphics[width=7cm]{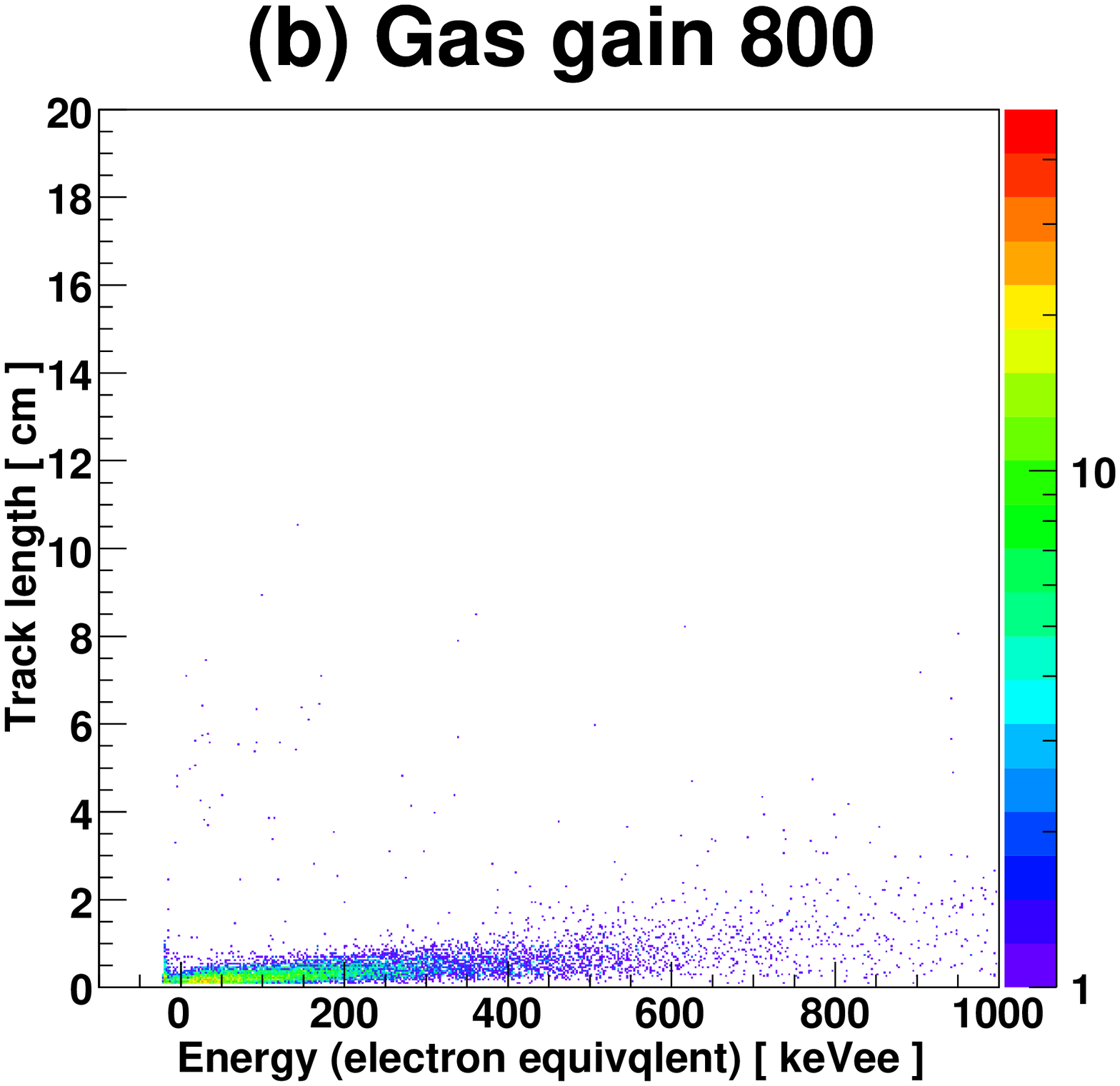}
\includegraphics[width=7cm]{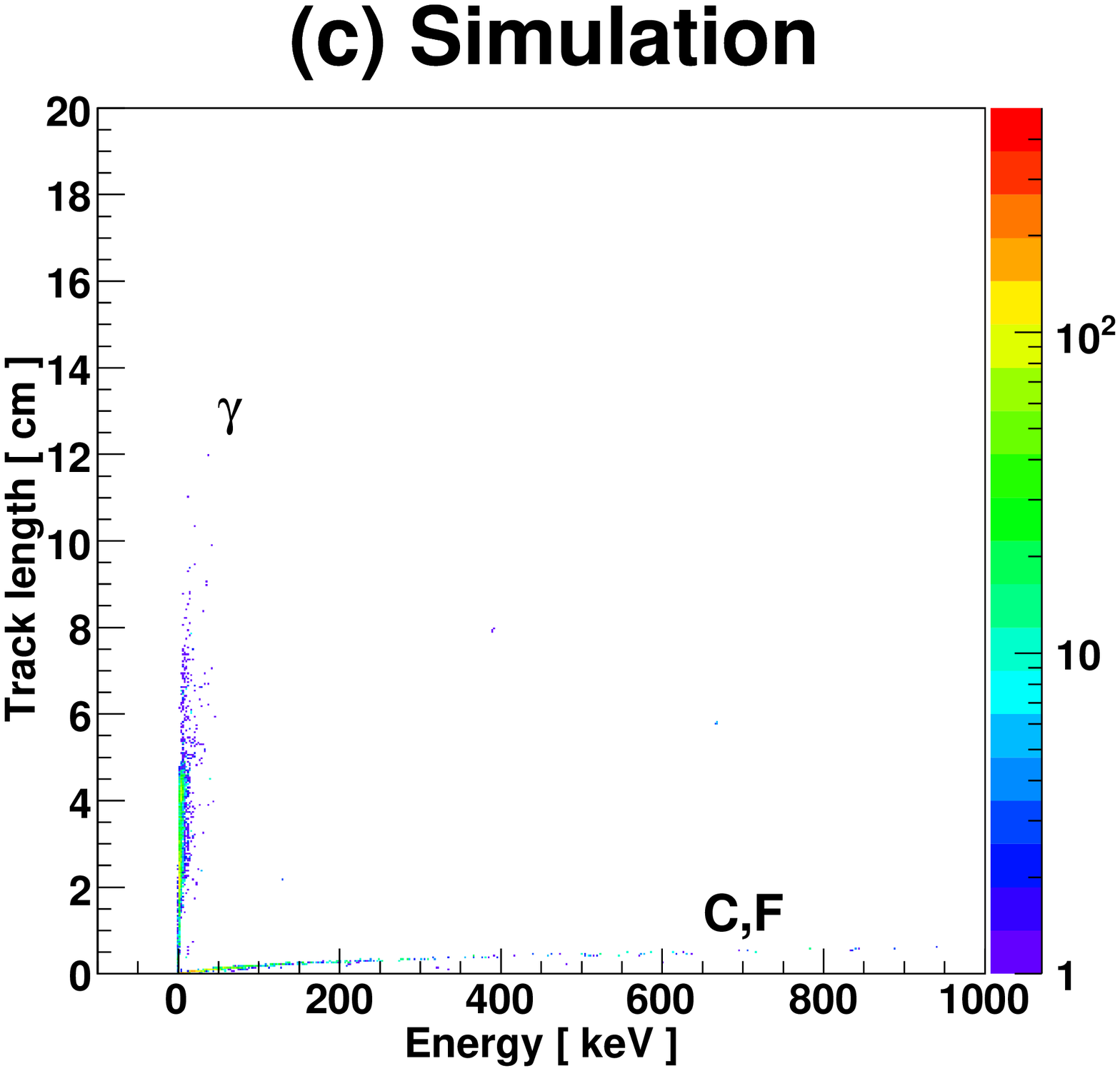}
\caption{Track length as a function of deposited energy with 150 Torr
 CF$_4$. (a) High-gain operation (low $dE/dx$ thresold),
(b) Low-gain operation (high $dE/dx$ thresold),
(c) Geant4 MC simulation without consideration of the diffusion of
drift process,  the energy resolution, and  the detection threshold.}
\label{fig:CF4TE}
\end{center}
\end{figure}

The measured track length of events  
when the GEM voltage was set to 215 V (gas gain of 4500) and 
95 V (gas gain of 800) are shown in Fig.\ref{fig:CF4TE}.
The threshold voltage of the discriminator of the ASD chip was
as high as 100mV; therefore, only C and F recoils were clearly
observed under operation with a gas gain of 800.

\section{Discussion and Prospects}
\begin{figure}[p]
\begin{center}
\includegraphics[width=7cm]{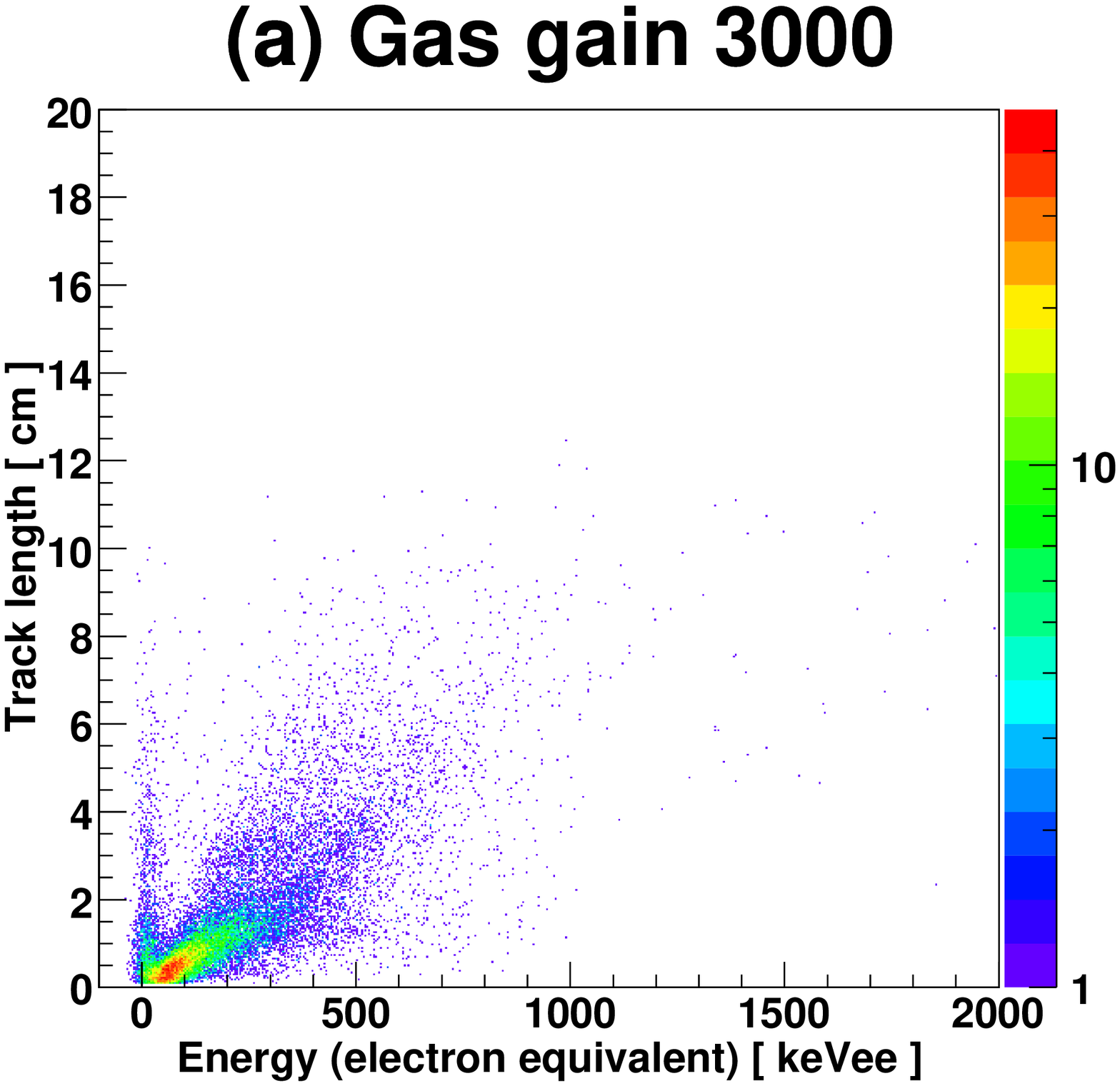}
\includegraphics[width=7cm]{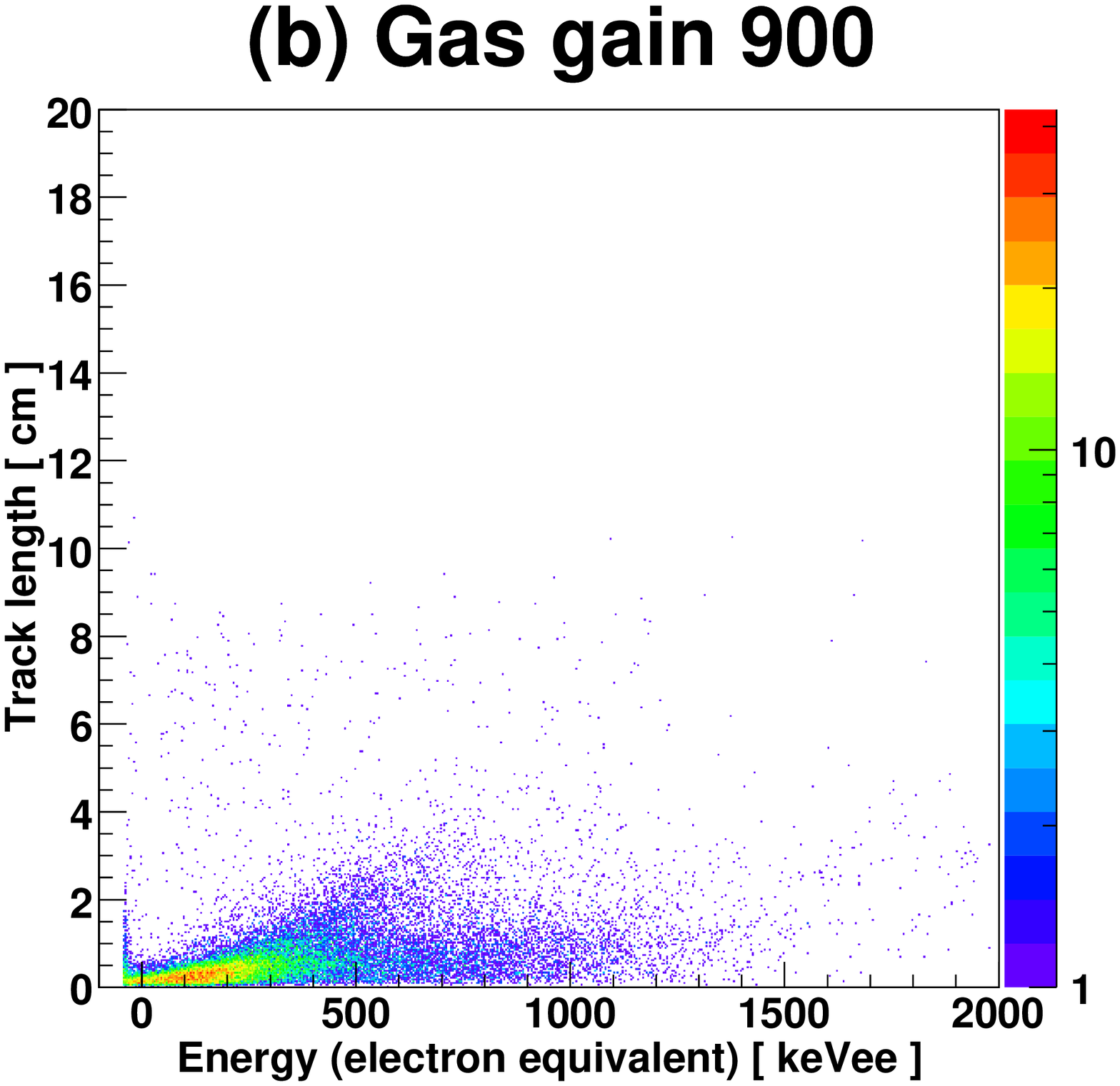}
\includegraphics[width=7cm]{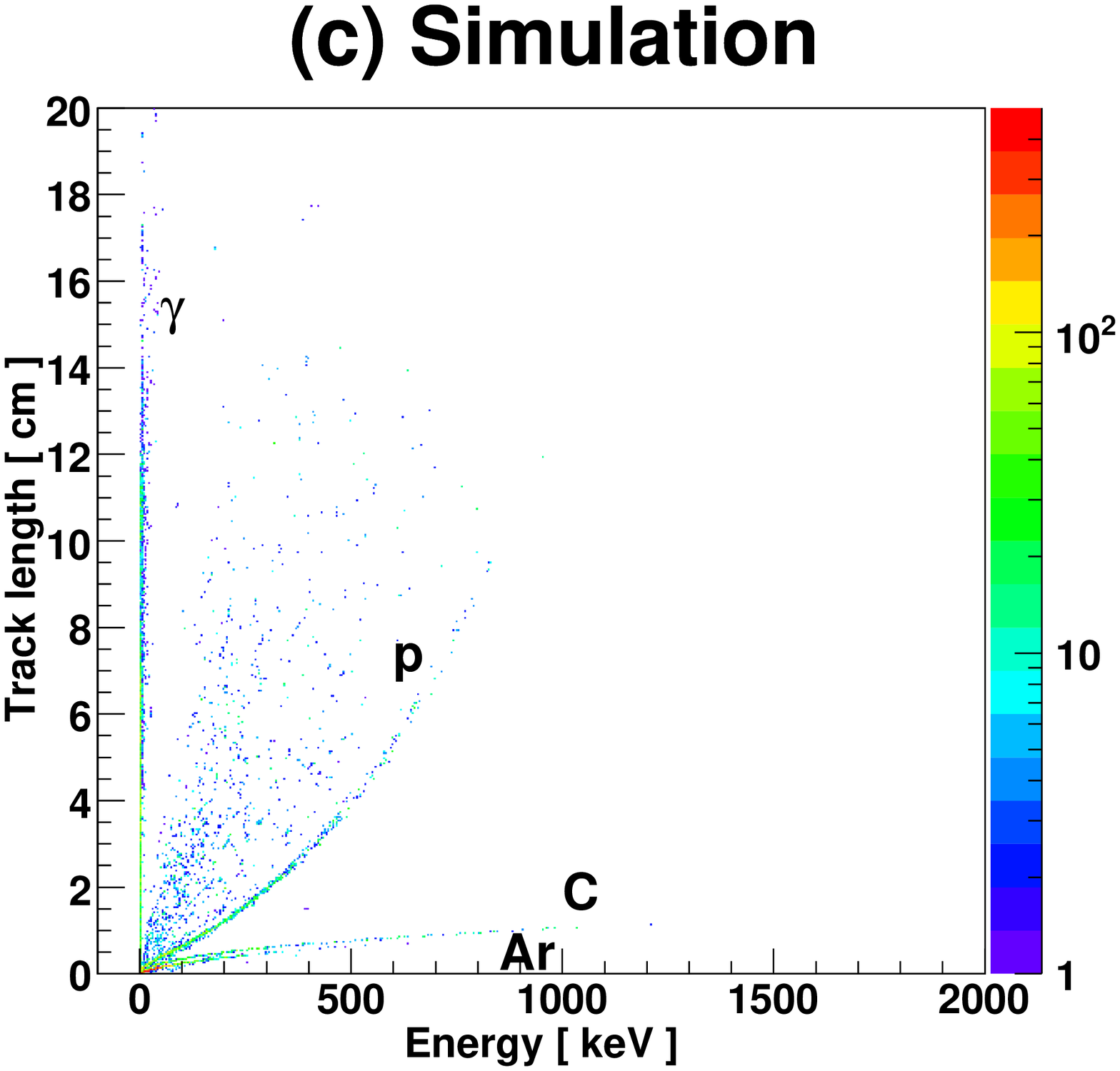}
\caption{Track length as a function of deposited energy with 150 Torr
 Ar-C$_2$H$_6$ (90:10). (a) High-gain operation (low $dE/dx$ thresold),
(b) Low-gain operation (high $dE/dx$ thresold),
(c) Geant4 MC simulation without consideration of the diffusion of
the drift process,  the energy resolution, and  the detection threshold.
Because the size of the TPC is too small for proton tracks,
proton events are scattered in upper regions of the main band. }
\label{fig:ArTE}
\end{center}
\end{figure}

We successfully showed the nuclear recoils
in 150 Torr of Ar-C$_2$H$_6$ (90:10) and CF$_4$ gases
according to their $dE/dx$ by changing the detector threshold.
The energy loss of protons
became lower as the energy increased as opposed to the other nuclei\cite{srim}.
Consequently,
the proton band in Fig. \ref{fig:ArTE}(b) is truncated
at the threshold set in the measurements,  
which corresponds to about 5 keV/400$\mu$m. 
In terms of $dE/dx$, the tracks in the micro TPC were much easier
to detect for C, F or Ar recoils.

Ultimately, our concern is 
the recoil direction of such nuclei below 100 keV 
to allow us to observe the signals of WIMPs.
In order to obtain longer tracks and clear Bragg curves, 
higher gas gain operations at lower pressures 
with low-energy neutron beams are needed.
The measurement of the incident neutron energy
with Time-Of-Flight may also be useful to examine the quenching factor
of nuclear ionization in the micro-TPC.



\end{document}